\documentclass[journal=jpccck,manuscript=article,layout=twocolumn]{achemso}
\usepackage[version=3]{mhchem}
\usepackage{balance}
\usepackage{times,mathptmx}
\usepackage{sectsty}
\usepackage{graphicx} 
\usepackage{float}
\usepackage[english]{babel}
\usepackage{array}
\usepackage[T1]{fontenc}
\usepackage[usenames,dvipsnames]{xcolor}
\usepackage{setspace}

\usepackage{subfigure}  
\usepackage{amsmath}

\author{Jaime Silva}
\affiliation{CFisUC, Department of Physics, University of Coimbra, Rua Larga, 3004-516 Coimbra, Portugal.}
\email{silva.jaime@gmail.com}
\phone{+351 239410681}
\author{Micael J.~T.~Oliveira}
\affiliation{Department of Physics, University of Li\`ege, B-4000, Li\`ege, Belgium}
\email{mjt.oliveira@ulg.ac.be}
\author{Senentxu Lanceros-Mendez}
\affiliation{Center of Physics, University of Minho, Campus de Gualtar, 4710-057 Braga, Portugal}
\alsoaffiliation{BCMaterials, Parque Cient\'ifico y Tecnol\'ogico de Bizkaia, 48160-Derio, Spain}
\alsoaffiliation{IKERBASQUE, Basque Foundation for Science, Bilbao, Spain}
\email{lanceros@fisica.uminho.pt}
\author{Fernando Nogueira}
\affiliation{CFisUC, Department of Physics, University of Coimbra, Rua Larga, 3004-516 Coimbra, Portugal.}
\email{fnog@uc.pt}
\phone{+351 239410114}

\title{Finite-size effects in the absorption spectra of a single-wall carbon nanotube}

\begin{document}

\begin{abstract}
The determination of the optical spectrum of single-wall carbon nanotubes (SWCNTs) is essential for the development of opto-electronic components and sensors with application in many fields. Real SWCNTs
are finite, but almost all the studies performed so far use infinite
SWCNTs. However, the spectra of finite and infinite systems are
different.  In this work the optical spectrum of finite (3,3) and
(5,5) SWCNTs is calculated as a function of nanotube length. For the
(3,3) SWCNTs, the calculated absorption spectra for light polarised
both parallel and perpendicularly to the nanotube axis are in good
agreement with experimental results. However, our
results indicate that the lowest energy peak present in the
experimental results for light polarised parallel to the nanotube axis
can be attributed to a surface-plasmon resonance that is a consequence
of the finite nature of the SWCNTs and not to the presence of SWCNTs
with other chiralities, as claimed by the previous theoretical works.
The surface-plasmon resonance is also studied using the Aharonov-Bohm effect.
Finally, this work demonstrates that the surface-plasmon resonance in finite SWCNT can be described using a 1D infinite well.
\end{abstract}

\section*{Introduction}

Historically, carbon nanotubes (CNTs) had their first appearance fifty
years ago as undesirable by-products of the fuel coke
industry~\cite{Monthioux2002}.  The subsequent detailed
characterisation presented in the seminal work of
Iijima~\cite{Iijima1991} for multiwall carbon nanotubes (MWCNTs) and,
in a later work, for single wall carbon nanotubes
(SWCNTs)~\cite{Iijima1993}, drew the scientific community's attention
to these carbon allotropes~\cite{Moniruzzaman2006,Charlier2007}. CNTs
can be considered as cylinders of covalently bonded carbon
atoms~\cite{Moniruzzaman2006} that, depending on how the graphene
honeycomb network is rolled up, can be either metallic or
semiconducting.  The application range for CNTs is vast and includes,
among others, conductive and high-strength composites, energy storage
and energy conversion devices~\cite{Baughman2002}. For CNTs, and in
particular for SWCNTs, the geometry of the SWCNT is related to how the
graphene honeycomb is rolled up and so the SWCNT surface can be
specified by a pair of integer numbers $(m,n)$ denoting the relative
position $\boldsymbol{C}_h=n\boldsymbol{a}_1 + m\boldsymbol{a}_2$ of a
pair of atoms in the graphene lattice, with $\boldsymbol{a}_1$ and
$\boldsymbol{a}_2$ being unit vectors of the hexagonal honeycomb
lattice~\cite{Charlier2007}.  An important question for both
metrological and applicative matters is how the optical absorption
spectrum varies with the CNT chirality as was recently demonstrated
experimentally by Vialla \textit{et al}~\cite{Vialla2013}.

Although, in general, CNTs available for experimental study are a
mixture of CNTs with different sizes and chiralities, the study of
straight SWCNTs with a well defined diameter was possible by using
$AIPO_4-5$ zeolite to align in their channels a SWCNT with a 4~\AA \,
diameter. Li \textit{et al}~\cite{Li2001} used this method to measure
the polarised absorption spectra of SWCNTs arrayed in the channels of
the $AIPO_4-5$ zeolite and drew the attention to the study of SWCNTs
with a well defined diameter but unknown or unspecified chirality.
The potential SWCNT candidates for such a small diameter are the
(3,3), (4,2) and (5,0) SWCNTs. Li \textit{et al}~\cite{Li2001}
reported that, when the exciting pulse electric field is polarised
parallel to the SWCNT axis, a sharp peak at 1.37~eV and two broad
bands at 2.1 and 3.1~eV are observed.  The same results were also
observed by Tang \textit{et al}~\cite{Tang2003}. It was also
demonstrated that the SWCNTs are transparent, due to a depolarisation
effect, for light polarised perpendicularly to the SWCNT direction.
The peaks' positions were explained through \textit{ab initio}
calculations using the plane-wave pseudo-potential
formulation~\cite{Kresse1993,Kresse1996} within the framework of the
Local Density Approximation (LDA) of Density-Functional Theory
(DFT)~\cite{Hohenberg1964,Kohn1965} and their origin is traced back to
transitions between bonding and antibonding $\pi$ orbitals.

Liu \textit{et al}~\cite{Liu2002}, using another local DFT
method\cite{Mintmire1983}, calculated the density of states (DOS), the
band structure and the absorption spectra of (3,3), (4,2) and (5,0)
SWCNTs oriented parallel to the light polarisation axis. They also
reported that the (5,0) and (3,3) SWCNTs are metallic while the (4,2)
is a small band gap semiconductor. The electronic properties, namely
the DOS and the band structure of (3,3), (4,2) and (5,0) SWCNTs, were
also addressed by Cabria \textit{et al}~\cite{Cabria2003}.

Mach\'{o}n \textit{et al}~\cite{Machon2002} performed LDA calculations
\cite{Sanchez1997,Soler2002} of the structural, electronic, and
optical properties of a 4~\AA \, SWCNT. When compared to the
experimental work of Li \textit{et al}~\cite{Li2001}, the calculated
optical properties revealed a discrepancy in the spectra for light
polarised perpendicularly to the SWCNT axis: the SWCNTs were not
transparent in the visible region. Li \textit{et al}~\cite{Li2001}
justified the disagreement by noting that the influence of the zeolite
on the optical properties had not been taken into account. This
disagreement of the theoretical and experimental optical absorption
spectra for light polarised perpendicularly to the SWCNT axis was
explained by Marinopoulos \textit{et
  al}~\cite{Marinopoulos2003}. These authors observed that theoretical
studies of the optical spectra of SWCNT had been neglecting crystal
local-field effects. By performing Time-Dependent DFT
(TDDFT)~\cite{Runge1984,marques2012fundamentals} calculations they
were able to reproduce the experimental spectra of Li \textit{et
al}~\cite{Li2001}. However, the CNTs considered were infinite (with 
different chiralities), being therefore different from the ones
used in the experiments.  In
this work we present a study of the absorption spectra of finite
SWCNTs of different lengths and further clarify the experimental
results of Li \textit{et al}~\cite{Li2001}.

\section*{Results and discussion}

As previously mentioned, TDDFT has already been successfully applied
to the calculation of the electronic excitations of SWCNTs.  In this
work we used the real-space code OCTOPUS~\cite{Andrade2012} to obtain
the TDDFT optical spectra of finite (3,3) and (5,5) SWCNTs in the
adiabatic TDLDA framework. The finite SWCNTs were
generated~\footnote{TubeGen 3.4 (web-interface,
  http://turin.nss.udel.edu/research/tubegenonline.html), J. T. Frey
  and D. J. Doren, University of Delaware, Newark DE, 2011} with
different $k$ translations of the unit cell along the tube
orientation, resulting in different SWCNT lengths as can be seen in Table \ref{tab1}.  

\begin {table} [ht]
  \begin{center}
    \begin{tabular}{c | c | c}
      $k$ & Length (\AA) & Length (\AA) \\ 
		    & SWCNT@(3,3) & SWCNT@(5,5)\\
	    \hline
		  5 & 13.22 & 13.28 \\
		  10 & 26.19 & 26.23 \\
		  15 & 39.00 & 39.19 \\
		  20 & 51.58 & 50.92 \\
		  25 & 63.90 & 63.00 \\
		  30 & 76.28 & 78.00 
    \end{tabular}
	\caption{\label{tab1} Table with with different used $k$ translations and associated SWCNT length}
  \end{center}
\end{table}

We used the default Troullier-Martins pseudopotentials~\cite{Troullier1991}
distributed with OCTOPUS and a uniform grid spacing of 0.20~\AA. The
simulation box was constructed joining spheres of radius
3.5~\AA\ around every atom. The time step used in the real-time
propagation of the time-dependent Kohn-Sham equations was 0.000658~fs,
ensuring the stability of the time propagation.  The terminations of
the finite SWCNT were passivated with hydrogens and the entire
structure was relaxed using a molecular dynamics code with the Merck
molecular force field (MMFF94)~\cite{Merck1996I,Merck1996II}.

For the infinite (3,3) SWCNT the code ABINIT~\cite{Gonze2009} was
used, also with the LDA approximation for exchange and correlation and
Troullier-Martins pseudo-potentials. The infinite SWCNT was placed
inside a parallelepipedic box of (25,25,2.464282)~\AA. The plane-wave
kinetic energy cutoff used was 35~Ha, with a k-point Monkhorst-Pack mesh
containing (1,1,512) points.

\begin{figure}[!ht]
  \centering
	\subfigure{
		\centering
		\includegraphics[width=0.45\textwidth]{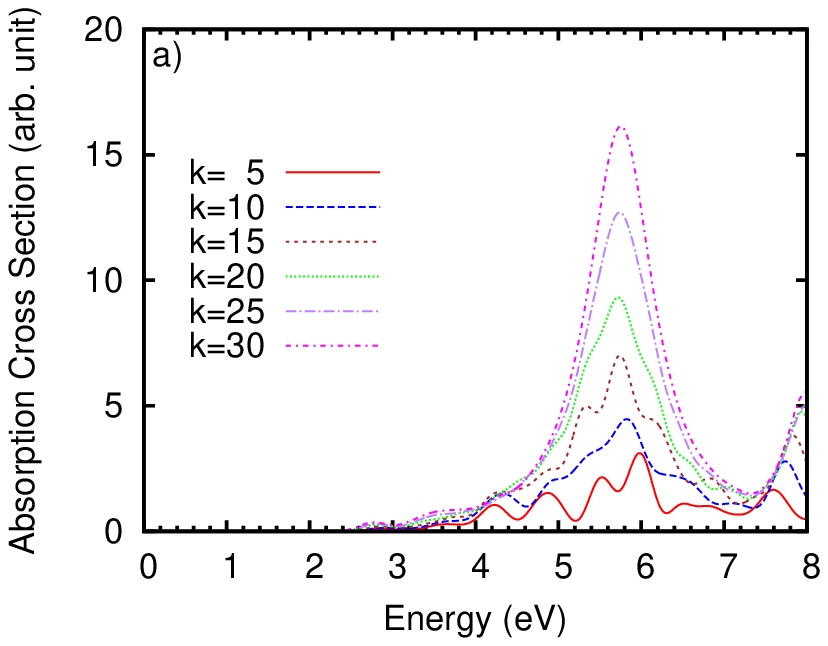}
	 }
	\subfigure{
	  \centering
		\includegraphics[width=0.45\textwidth]{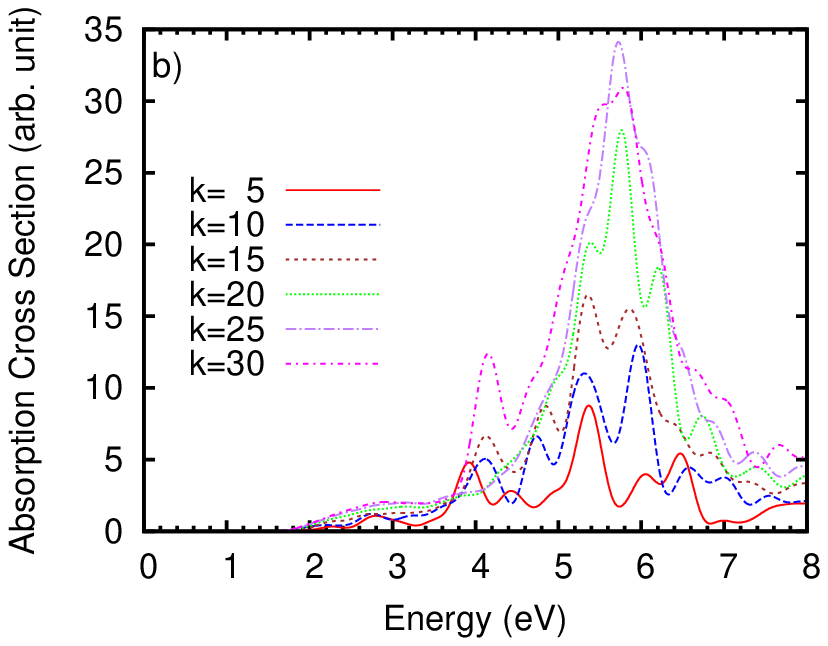}
	}
	\caption{\label{fig1}Absorption spectra of finite (3,3) (a)) and (5,5) (b)) SWCNTs of different lengths for incident light polarised perpendicularly to the SWCNT axis.}
\end{figure}

The absorption spectra for incident light polarised perpendicularly to
the (3,3) and (5,5) SWCNT axis for finite SWCNTs of different lengths
were calculated and the results are shown in Fig.~\ref{fig1}.
The spectra are essentially the same, showing that the SWCNTs are
transparent until around 4 eV, in good agreement with the experimental
work of Li \textit{et al} \cite{Li2001} and Tang \textit{et
  al}~\cite{Tang2003} for the (3,3) SWCNTs. The position of the first
large peak is also in agreement with the results of the TDDFT
calculations of Marinopoulos \textit{et al}~\cite{Marinopoulos2003}
for the (3,3) SWCNTs. The increase in the height of the peaks with
increasing SWCNT length is simply a reflection of the increase in the
number of electrons that are available to be excited due to the
increase in the number of carbon atoms.

\begin{figure}[!ht]
  \centering
	\subfigure{
	  \centering
		\includegraphics[width=0.45\textwidth]{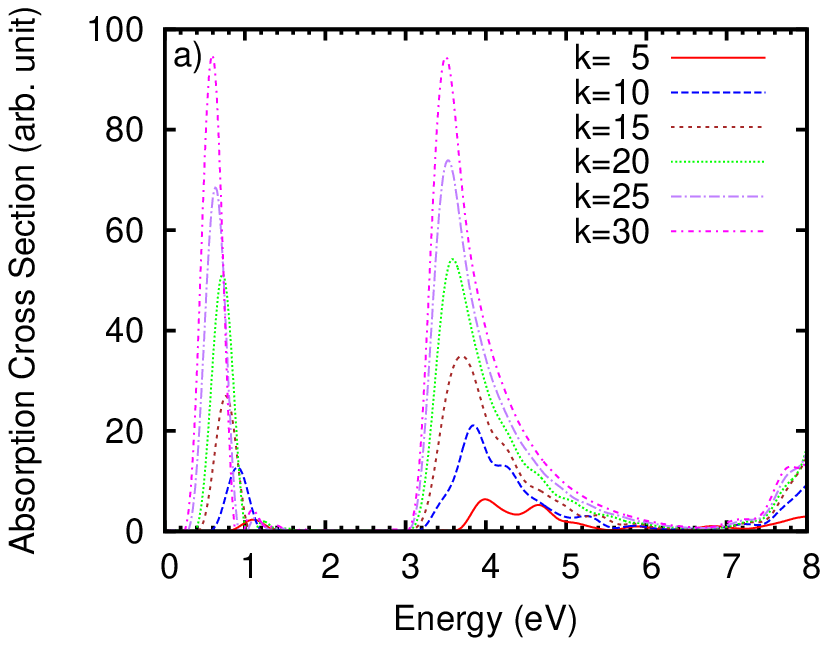}
	}
	\subfigure{
	  \centering
		\includegraphics[width=0.45\textwidth]{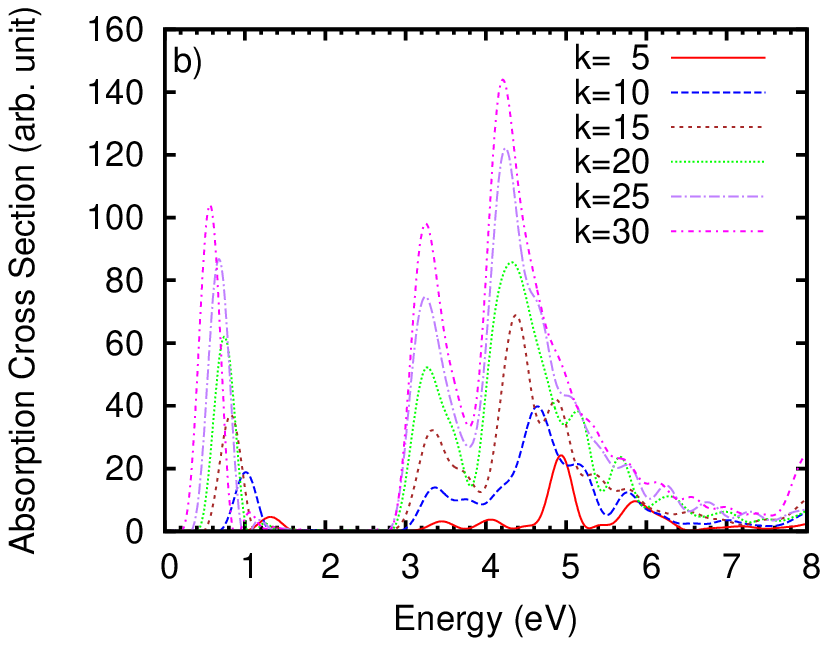}
	}
	\caption{\label{fig2}Absorption spectra of finite (3,3) (a)) and (5,5) (b)) SWCNTs of different lengths for incident light polarised parallel to the SWCNT axis.}
\end{figure}

In Fig.~\ref{fig2} the absorption spectra for light polarised parallel
to the above mentioned (3,3) and (5,5) SWCNT axes are shown. For the
(3,3) SWCNTs there is a peak between $1.11$ and $0.59$~eV, depending
on nanotube length, a second peak between $3.99$ and $3.5$~eV. 
As in the case of perpendicularly polarised light, the height of 
the peaks increases with increasing SWCNT length. 
Furthermore, the first two peaks can also be observed
for the (5,5) SWCNT at similar energies (and a third one at higher energies).

It is apparent from Fig.~\ref{fig2} that increasing the length of the
SWCNT redshifts the lowest peaks, as expected for finite systems. This
can be traced back to the fact that an approximated $f_\mathrm{xc}$
kernel (as the TDLDA used in this work) neglects dynamical
correlations effects~\cite{Onida2002}. Nevertheless, it can be also
attributed to a surface-plasmon resonance as it can be observed for
simple atom chains~\cite{Yan2008}. Later in the text it is
demonstrated that the first peak can indeed be attributed to a
surface-plasmon resonance.

A notable difference between this work and previous ones is that the
first peak observed in Fig.~\ref{fig2} is not present in the infinite
(3,3) SWCNT calculations as, \textit{e.g.}, those of Marinopoulos
\textit{et al}~\cite{Marinopoulos2003} and Tang \textit{et al}
\cite{Tang2003}, although it does appear in the experimental results
of Li \textit{et al}~\cite{Li2001} and Tang \textit{et al}
\cite{Tang2003}. In previous theoretical/computational works the
presence of this peak was explained assuming the presence of SWCNTs
with different chiralities, namely the (3,3), (4,2), and (5,0)
SWCNTs. Our results suggest that the absence of the peak in previous
calculations is related to the infinite nature of the SWCNTs used. Our
finite SWCNTs calculations do show that the peak should be present in
samples consisting exclusively of $(n,n)$ SWCNTs like the (3,3) and
(5,5) nanotubes. This claim would be impossible to check without a way
of preparing SWCNT samples with a single chirality. Fortunately, the
recent work of Blancon \textit{et al}~\cite{Blancon2013} opened up
this possibility, allowing our claim to be verified.

\begin{figure}[!ht]
  \centering
		\includegraphics[width=0.45\textwidth]{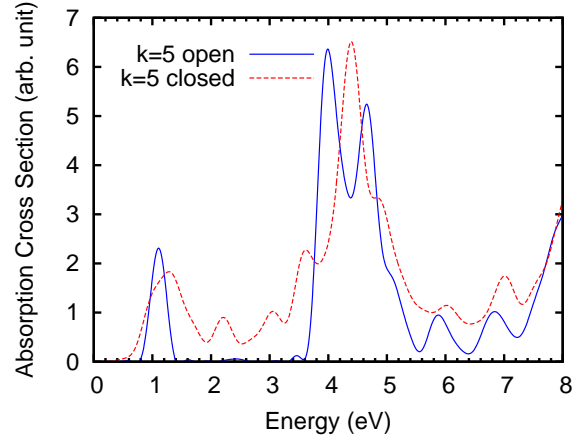} 
		\caption{\label{fig3} Absorption spectra of finite (3,3) SWCNTs of length $k=5$ for incident light polarised parallel to the SWCNT axis with open and closed ends, terminated with carbon atoms.}
\end{figure}

One straight explanation for the origin of the first peak observed in
Fig.~\ref{fig2}, is that it can be due to states localised at the ends
of the SWCNT. To demonstrate the latter possibility,
the optical spectra presented in Fig.~\ref{fig3}, for incident light
polarised parallel to the SWCNT, was calculated for the (3,3) SWCNT with $k = 5$ with
closed ends (the open-ends SWCNT was terminated with carbon atoms), using the same methodology
as described previously.

\begin{figure}[!ht]
  \centering
		\includegraphics[width=0.45\textwidth]{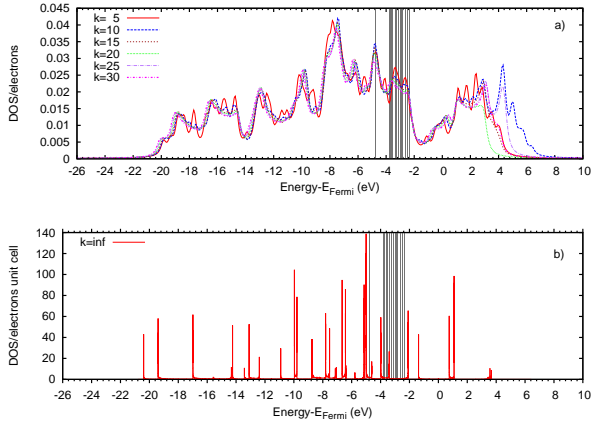} 
		\caption{\label{fig4} a) Normalised DOS for finite SWCNTs and b) for an infinite SWCNT. The 
		grey vertical lines indicate the states that contribute to the first peak of the absorption spectra 
		of the finite SWCNTs. The Fermi level was set at 0~eV in both cases.}
\end{figure}

As can be seen in Fig.~\ref{fig3}, when comparing open and closed ends SWCNTs, the low
energy peak is in the same position, but is broader for the closed SWCNT. This is due to the additional set of carbon atoms at both ends. Therefore, the
origin of this peak can not be transitions involving localised states
near the ends of the SWCNT.
But the logical explanation for the absence of these peaks
from the infinite SWCNT calculations is that the Kohn-Sham states involved in 
this low-energy excitation
are present in the finite SWCNTs but absent in the infinite
one. Due to the size of the systems, the number of states present in
them is too large to allow for an easy understanding of the origin of
the peak. We therefore decided to look for the relevant states putting
them in resonance with an applied laser. We studied the
effect of an incident laser, polarised parallel to the SWCNT axis,
with a frequency equal to the first peak of the (3,3) SWCNT with $k =
20$ (0.79~eV), a pulse width of 32.91~fs and an enveloping cosine
function. This study was performed with the OCTOPUS code for all the
finite SWCNTs considered above. The resonant states were found by
plotting the probability density for each state, after the application
of the laser, and observing which states were disturbed by it.
Combining this information with the density of states (DOS) of each
nanotube readily highlights the difference between the finite and
infinite nanotubes, as can be seen in Fig.~\ref{fig4}.
In this figure, the DOS for the different finite SWCNT lengths was
normalised to the number of electrons of each nanotube. The grey
vertical lines indicate the resonant states found with the above
described procedure (with a linewidth of 0.04 eV). The analysis of this figure clearly shows that
there are states which are present in the finite nanotubes but
disappear when the nanotube becomes infinite. These states are the
ones contributing to the lowest energy peak, present in our
calculations and on the experimental results of Li \textit{et al}
\cite{Li2001}, but not observed in the infinite nanotube calculations
of Refs.~\cite{Marinopoulos2003} and \cite{Tang2003}. Note that the
large differences above 3~eV in Fig.~\ref{fig4}a) are simply due to
the different number of unoccupied states considered in the
calculations. Moreover, this nanotube is metallic and exhibits states 
around the Fermi level. These states are difficult to observe in the 
plot due to the large number of van Hove singularities.

\begin{figure}[!ht]
  \centering
		\includegraphics[width=0.45\textwidth]{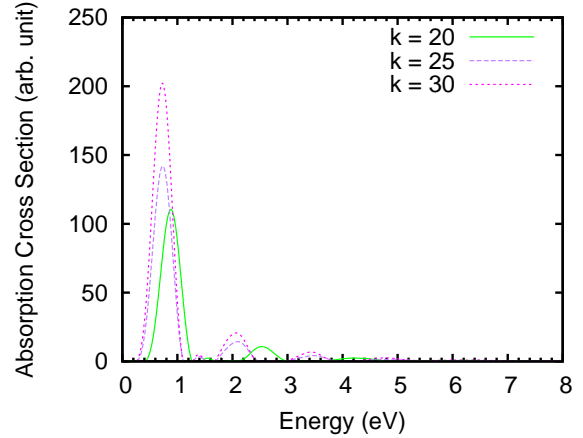}
		\caption{\label{fig5} Absorption spectra for a set of independent particles in a 1D infinite well with length equal to the (3,3) SWCNT}
\end{figure}

From the results presented previously it is possible to state that
this first peak is a surface-plasmon resonance and that its origin is
the finite nature of the SWCNT. The confinement of the wave function
on the finite 1D SWCNT leads to a standing wave
and that shows up in the spectra as a surface-plasmon resonance peak.
This can be replicated with a set of independent particles that are
confined in a 1D infinite well that has a length equal to the SWCNT.
In fact, Fig. \ref{fig5} presents the absorption spectra for
a set of independent particles, $N=20,70,100$, in a 1D infinite well
with length equal to the SWCNTs with, respectively, $k=20,25,30$. 
The TDDFT calculation was done using the exact same parameters, although
the electron-electron interaction terms in the Hamiltonian 
(Hartree, exchange and correlation potentials) were set to zero.

By comparing Fig.~\ref{fig5} with Fig.~\ref{fig2} it is possible to
observe that the positions of the first peaks are
similar.
Furthermore, the origin of the small peaks that appear when the length
of the SWCNT increases can be explained by the dependence of the
single particle transition energies on the length of the 1D infinite
well or SWCNT.
When the SWCNT length increases the energy of the single particle
transitions decreases, appearing as a peak at lower energies.
Moreover, it was predicted by Nakanishi \textit{et al}~\cite{Nakanishi2009} that,
at the far infrared regime, the absorption spectra should exhibit a surface plasmon resonance peak.
The authors of this work use a self consistent method that depends on an initial, 
frequency dependent, incident electric field and on the non-local conductivity of the nanotube.
The predicted far infrared surface plasmon resonance peak was 
recently observed experimentally~\cite{Morimoto2014} for SWCNTs that have a length of the order of 1~$\mu$m.
We stress that in our work the origin of the surface plasmon is distinctly different from the 
original proposal by Nakanishi \textit{et al}~\cite{Nakanishi2009}, i. e., in our work the 
calculations are \textit{ab-initio} and do not depend on a particular electric field wavelength.

\begin{figure}[!ht]
  \centering
		\includegraphics[width=0.45\textwidth]{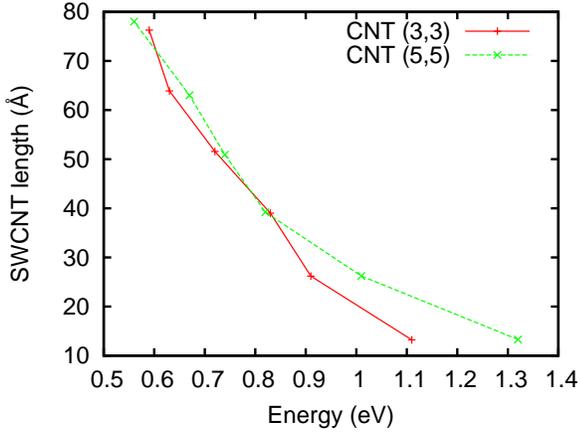}
		\caption{\label{fig6} Lenght of the SWCNT versus the energy of the first peak for two types of armchair SWCNT.}
\end{figure}

This lower energy peak has been observed in previous
experimental work.
The experimental spectra of Li \textit{et al}~\cite{Li2001} and Tang
\textit{et al}~\cite{Tang2003} for light polarised parallel to the
nanotube axis show a peak at 1.37 eV and two broad bands centred at
2.1 and 3.1 eV. Comparing these results to those presented in
Fig.~\ref{fig2}, it is apparent that the first peak of Fig.~\ref{fig2}
is redshifted with respect to the first experimental peak and the
second peak is blue-shifted relative to the third experimental peak.
The broad experimental peak centred at 2.1 eV, however, is not
present in our results. 
Previous attempts to theoretically reproduce the experimental
results, like the work of Marinopoulos \textit{et al}~\cite{Marinopoulos2003},
Machon \textit{et al}~\cite{Machon2002} and Spataru \textit{et
  al}~\cite{Spataru2004}, required the conjugation of the
absorption spectra of three SWCNT with different chiralities.
Moreover, as demonstrated by Spataru \textit{et al}~\cite{Spataru2004} for the (3,3) SWCNT,
there is only one bound exciton that positions the third peak of the (3,3) absorption spectra at 3.17 eV.

Another important question is how the energy of the first peak evolves with the increase in length of the SWCNT. 
Fig.~\ref{fig6} is a plot of the energy of the first peak versus the length of the tube for the two types of 
SWCNT, (3,3) and (5,5).

It is possible to observe that for the low aspect ratio (3,3) and (5,5) SWCNTs the energy of the first peaks of 
the (3,3) tubes differs from that of the (5,5) tubes. 
The latter is due to curvature effects (curvature is so strong that some rehybridization among the sigma and pi states appears) that alters the electronic structure of the nanotube and is more pronounced small diameter SWCNT \cite{Charlier2007} like the (3,3) SWCNT.
Nevertheless, with the increase of SWCNT length, the energies of the first peaks are similar for the two types of 
armchair SWCNT, which is consistent with the 1D infinite well approximation.
Moreover, the results presented in Fig.~\ref{fig6} do not follow a linear relation
and, as expected for a surface plasmon resonance peak,
the position of the peak depends on the length of the SWCNT. 
The presence of surface plasmon peak at low energies enables the use of SWCNT as 
plasmonics-active systems \cite{Vo-Dinh2010}.
The application of the surface plasmon resonance is vast and can be use in bioanalytic 
and biophysical applications \cite{Couture2013}.
In fact for SWCNT there is increase interest in studying the influence of the surface 
plasmon resonance with SWCNT physical properties as is demonstrated, for higher frequency
plasmon, in this recent work \cite{Rance2010}.
Finally, it is important to point that TDDFT was employed to study the surface plasmon
resonance of other systems \cite{Lopez-Lozano2014}

It is well known that, when modelling a finite portion of an infinite metallic nanotube,
the HOMO-LUMO gap usually opens. Low-energy peaks could be attributed to this gap opening. However,
in this study the gap is very small and this explanation for the low-energy peaks can be
safely discarded.

\begin{figure}[!ht]
  \centering
		\includegraphics[width=0.45\textwidth]{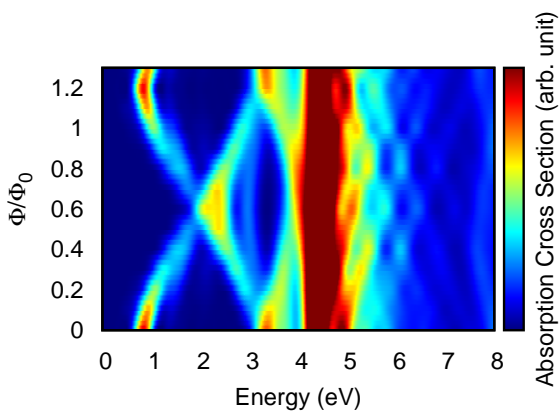}
		\caption{\label{fig7} Absorption cross section versus applied magnetic flux for the (5,5) CNT.}
\end{figure}

Nevertheless, one can test the effect of the HOMO-LUMO gap opening in the absorption spectra of a finite (5,5) CNT using the Aharonov-Bohm effect\cite{Aharonov1961}.
For that we add to the normal DFT/TDFT calculations a static magnetic field parallel to the CNT main axes with a value, for the (5,5) CNT with $k=15$, between $1043.4$ and $1.3564 \times 10^4$ Tesla.
In a celebrated work,\cite{Aharonov1961} Aharonov and Bohm describe a possible experiment to study the quantum effects of the application of a potential to charged particles.
In their work\cite{Aharonov1961} it is described an effect that arises from the quantum nature of the matter, i. e., a charged particle feels the effect of a potential that was acting on it even if there is no field in the current region.
To explain the latter effect, Aharonov and Bohm proposed a hypothetical experiment where it is demonstrated that, by using a magnetic flux in a selected region of space, two electron beams acquire a phase difference that is proportional to the magnetic flux quantum, $\phi_0=\frac{h}{e}$ with $h$ as the Planck constant and $e$ the electronic charge, even if the magnetic flux has a zero value in the region where the measurement is performed. 
The Aharonov-Bohm effect\cite{Aharonov1961} was demonstrated experimentally by Tonomura et al.\cite{Tonomura1986} in 1986 starting the application and study of this effect to meso and nano systems.
For infinite CNT, the application of the Aharonov-Bohm effect results in oscillations of the band gap first described by Ajiki and Ando \cite{Ajiki1993,Ajiki1994}.
Furthermore, in the case of the metallic CNT, there is a gap opening that is dependent on the applied magnetic flux and has $\phi_0$ periodicity\cite{Ajiki1993,Ajiki1994}.
More recently it has been shown that curvature effects break the periodicity of the gap oscillation \cite{Sangalli2011}.  
In Fig.~\ref{fig7} it is presented the absorption cross section versus applied magnetic flux for the (5,5) CNT. 
It is possible to observe the existence of two type of peaks. The first one, the plasmonic peak, oscillating with the increase of the applied magnetic flux, and the usual metallic peak splitting with the application of magnetic flux. 

\begin{figure}[!ht]
  \centering
		\includegraphics[width=0.45\textwidth]{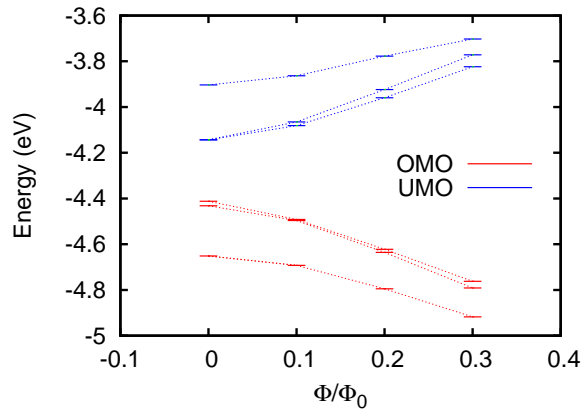}
		\caption{\label{fig8} Occupied molecular orbitals (OMO) and unoccupied molecular orbitals (UMO) energies versus magnetic flux for the (5,5) CNT, showing the HOMO and LUMO followed by the two immediately molecular orbitals.}
\end{figure}

As previously explained, with the increase of the magnetic flux the band gap oscillates and, for finite nanotubes, the energy difference between the highest occupied molecular orbital (HOMO) and the lowest unoccupied molecular orbital (LUMO) also oscillates with a periodicity higher than $\phi_0$, as can be seen in Fig.~\ref{fig8}
This oscillation of the HOMO - LUMO gap is at the origin of the plasmonic peak position with the increase of the magnetic flux. This is due to the fact that the peak is only related with independent particles transitions in a confined environment and so it depends on the energy difference between the occupied an unoccupied orbitals.
In the same figure it is possible to observe that the metallic peak splits with the increase of the magnetic flux.
The splitting can be explained for infinite CNT with the lifting of the van Hove singularities \cite{Charlier2007,Roche2000}.
For the finite CNT the metallic peaks spliting is related to the lifting of degenerated molecular orbitals, as can be seen observed in figure Fig.~\ref{fig8}.

\begin{figure}[!ht]
  \centering
		\includegraphics[width=0.45\textwidth]{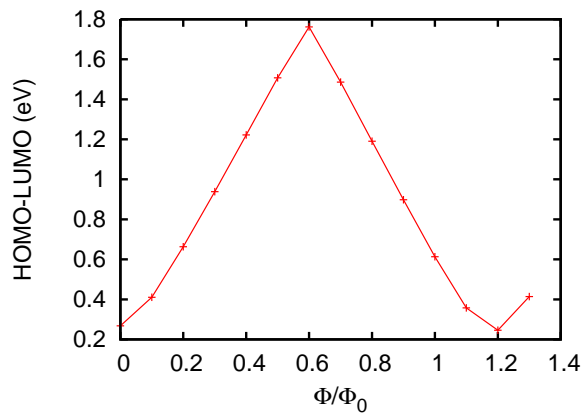}
		\caption{\label{fig9} HOMO-LUMO difference versus magnetic flux for the (5,5) CNT.}
\end{figure}

It can also be seen in Fig.~\ref{fig7} that the surface plasmon peak does not split, hence the origin of the plasmon peak is not related to transitions from/to degenerated molecular orbitals or, in the case of infinite CNT, from/to the van Hove singularities, as demonstrated previously.
Another point, is the periodicity of the HOMO-LUMO gap presented in Fig.~\ref{fig9}, as it can be seen that the period is greater than $\phi_0$. 

This anomalous Aharonov-Bohm effect, as explained by Sangalli et. al. \cite{Sangalli2011}, is related to curvature effects or to impurities and defects that creates deviations from a perfect circle of radius $R_{CNT}$. 
In fact, by employing an optimization procedure to relax the finite CNT structures produces oscillations on the value of $R_{CNT}$, mainly at the ends of the CNT.

\section*{Conclusions}

In this work the optical spectra of finite (3,3) and (5,5) SWCNTs with different lengths were
calculated. It was found that these
nanotubes are transparent to visible and near-UV light polarised
perpendicularly to their axis, in agreement with previous experimental
and theoretical results. It was also shown that the absorption spectra
for light polarised parallel to the nanotubes' axes exhibit a low energy 
peak that is redshifted with the increase of the SWCNT length. This peak is
present in previous experimental studies, and
our results suggest that its absence from previous
calculations is related to the infinite nature of the SWCNTs used in those and
not to the presence of nanotubes with other chiralities, as previously thought.
The method recently proposed by Blancon \textit{et al}~\cite{Blancon2013}
will allow for the validation of this claim.
This low energy peak is studied using the Aharonov-Bohm effect in CNT confirming that it is a plasmonic peak energy and is due to the finite nature of CNT and also that it can be described using 1D infinite well. 
Finally the relation between the plasmonic peak position and the SWCNT length is also a signature 1D infinite well. 
Nevertheless, for the longer SWCNT (longer than the SWCNT study in this work) other effects must be considered like the van der Waals interaction.

\section*{Acknowledgments}
We thank the Laboratory for Advanced Computing (LCA) of the University of Coimbra
for the computer time allocated to this project.

\balance

\bibliography{references} 

\clearpage
\onecolumn
\section*{TOC}
\begin{figure}[!ht]
  \centering
		\includegraphics[width=1.0\textwidth]{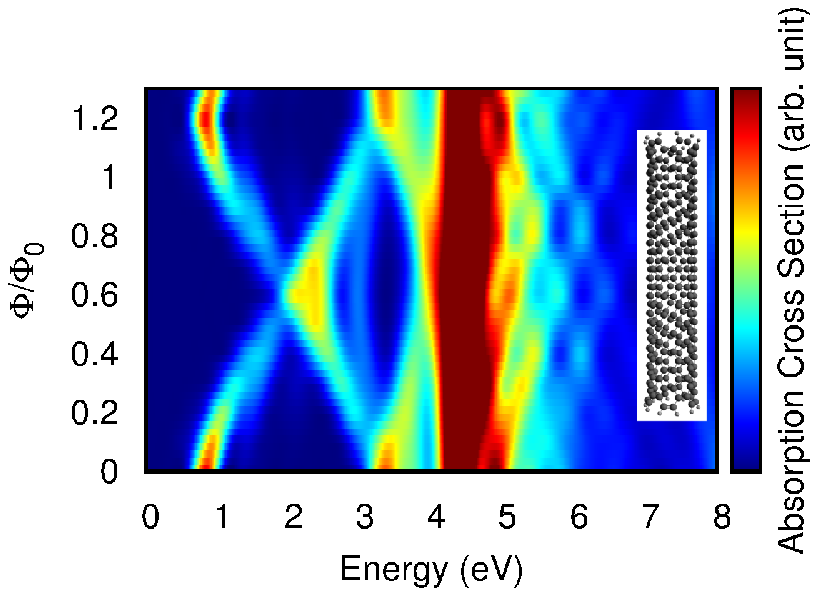}
		\caption{\label{toc} TOC.}
\end{figure}

\end{document}